\documentclass[%
aip,
amsmath,amssymb,
reprint,%
]{revtex4-1}

\usepackage{graphicx}
\usepackage{dcolumn}
\usepackage{bm}
\usepackage[english]{isodate}
\usepackage[dvipsnames]{xcolor}
\usepackage{lipsum}
\usepackage[utf8]{inputenc}
\usepackage[T1]{fontenc}
\usepackage{mathptmx}
\usepackage{etoolbox}
\usepackage{tabularx}
\usepackage{xcolor}
\usepackage{chemformula}
\usepackage{siunitx}
\usepackage{booktabs}

\newcommand{\green}[1]{{\color{Black}{#1}}}%
\makeatletter
\def\@email#1#2{%
	\endgroup
	\patchcmd{\titleblock@produce}
	{\frontmatter@RRAPformat}
	{\frontmatter@RRAPformat{\produce@RRAP{*#1\href{mailto:#2}{#2}}}\frontmatter@RRAPformat}
	{}{}
}%
\makeatother
\begin{document}
	\preprint{AIP/123-QED}
	
	\title{Crystallization dynamics of amorphous yttrium iron garnet thin films}
	\author{Sebastian Sailler} 
	\email{sebastian.sailler@uni-konstanz.de}
	\affiliation{Department of Physics, University of Konstanz, 78457 Konstanz, Germany} 
	
	\author{Gregor Skobjin} 
	\affiliation{Department of Physics, University of Konstanz, 78457 Konstanz, Germany}
	
	\author{Heike Schl\"orb} 
	\affiliation{Leibniz Institute of Solid State and Materials Science, 01069 Dresden, Germany}
	
	\author{Benny Boehm} 
	\affiliation{Institute of Physics, Technische Universit\"at Chemnitz, 09126 Chemnitz, Germany}
	
	\author{Olav Hellwig} 
	\affiliation{Institute of Physics, Technische Universit\"at Chemnitz, 09126 Chemnitz, Germany} 
	\affiliation{Center for Materials Architectures and Integration of Nanomembranes (MAIN), Technische Universit\"at Chemnitz, 09107 Chemnitz, Germany}

	\author{Andy Thomas} \affiliation{Leibniz Institute of Solid State and Materials Science, 01069 Dresden, Germany}
	\affiliation{Institut f\"ur Festk\"orper- und Materialphysik (IFMP), TUD Dresden University of Technology, 01069 Dresden, Germany}
	
	\author{Sebastian T. B. Goennenwein} 
	\affiliation{Department of Physics, University of Konstanz, 78457 Konstanz, Germany}
	
	\author{Michaela Lammel} 
	\email{michaela.lammel@uni-konstanz.de}
	\affiliation{Department of Physics, University of Konstanz, 78457 Konstanz, Germany}

	\date{\today}
	
	\begin{abstract}
		Yttrium iron garnet (YIG) is a prototypical material in spintronics due to its exceptional magnetic properties. To exploit these properties high quality thin films need to be manufactured. Deposition techniques like sputter deposition or pulsed laser deposition at ambient temperature produce amorphous films, which need a post annealing step to induce crystallization. However, not much is known about the exact dynamics of the formation of crystalline YIG out of the amorphous phase. Here, we conduct extensive time and temperature series to study the crystallization behavior of YIG on various substrates and extract the crystallization velocities as well as the activation energies needed to promote crystallization. We find that the type of crystallization as well as the crystallization velocity depend on the lattice mismatch to the substrate. We compare the crystallization parameters found in literature with our results and find an excellent agreement with our model. Our results allow us to determine the time needed for the formation of a fully crystalline film of arbitrary thickness for any temperature. 
	\end{abstract}
	
	\maketitle
	
	\section{Introduction}
	Yttrium iron garnet (\ch{Y3Fe5O12}, YIG) is an electrically insulating ferrimagnet, crystallizing in a cubic crystal lattice with Ia$\bar{3}$d symmetry. \cite{bertaut1956structure, geller_crystal_1957} Its electric and magnetic properties include a long spin diffusion length, which makes YIG an ideal material for spin transport experiments with pure spin currents. \cite{kajiwara_transmission_2010, althammer_quantitative_2013, cornelissen2015long} Additionally, YIG shows an exceptionally low Gilbert damping and a low coercive field, which allows investigations of magnon dynamics via e.g. ferromagnetic resonance experiments. \cite{dillon_1957_FMR, cherepanov_saga_1993, Wang_Large_spin_pumping, sakimura_nonlinear_2014, dubs_low_2020} These exceptional properties caused YIG to be intensively studied and made it the prototypical material in the field of spintronics, which almost exclusively relies on devices in thin film geometry. 
	
	\green{Several deposition techniques are known to produce high quality YIG thin films, including pulsed laser deposition (PLD), \cite{heinrich2006pulsed, krockenberger_solid_2008, haidar_thickness-_2015, hauser_yttrium_2016, hauser_annealing_2017, heyroth_monocrystalline_2019, gurjar_control_2021, dallivy_kelly_inverse_2013} liquid phase epitaxy (LPE) \cite{shone_technology_1985, gornert_growth_1988, cermak_yig_1990, beaulieu_temperature_2018, dubs_low_2020, will-cole_negligible_2023} and radio-frequency (RF) magnetron sputtering. \cite{park_structural_2001, jang_new_2001, boudiar_magneto-optical_2004, yamamoto_post-annealing_2004, kang_magnetic_2005, block_growth_2014, liu_ferromagnetic_2014, chang_nanometer-thick_2014, lustikova_spin_2014, jungfleisch_spin_2015, zhang_growth_2015, li_epitaxial_2016, cooper_unexpected_2017, lian_annealing_2017, talalaevskij_magnetic_2017, zhu_patterned_2017, ding_nanometer-thick_2020, ding_sputtering_2020}} Some deposition techniques like magnetron sputtering give the opportunity to deposit both, amorphous and crystalline thin films, depending on the process temperatures during deposition. \cite{jang_new_2001, jang_growth_2004} Here, room temperature magnetron sputtering processes yield amorphous films. \cite{park_structural_2001, jang_new_2001, boudiar_magneto-optical_2004, yamamoto_post-annealing_2004, kang_magnetic_2005, block_growth_2014, liu_ferromagnetic_2014, chang_nanometer-thick_2014, lustikova_spin_2014, jungfleisch_spin_2015, li_epitaxial_2016, cooper_unexpected_2017, lian_annealing_2017, talalaevskij_magnetic_2017, zhu_patterned_2017, ding_nanometer-thick_2020, ding_sputtering_2020, jang_growth_2004} For the deposition of YIG onto gadolinium gallium garnet (\ch{Gd3Ga5O12}, GGG) substrates, which feature a lattice constant very similar to the one of YIG, direct epitaxial growth was observed for process temperatures of \SI{700}{\degreeCelsius}. \cite{jang_new_2001, jang_growth_2004} On quartz a post annealing step is needed to enable the formation of polycrystalline YIG. \cite{roumie_effect_2010}
	
	\green{The annealing process is usually performed in air \cite{boudiar_magneto-optical_2004, cooper_unexpected_2017} or reduced oxygen atmosphere \cite{kang_magnetic_2005, bai_characterization_2019, ding_nanometer-thick_2020, seol_development_2023} to counteract potential oxygen vacancies in the YIG lattice. For amorphous PLD films annealing in inert argon atmosphere has been reported to have no deteriorating influence. \cite{hauser_annealing_2017} Annealing crystalline, sputtered YIG films in vacuum, however, showed a reduction in typical characteristic properties like the spin Hall magnetoresistance in YIG/Pt. \cite{bai_characterization_2019}} 
	
	\green{Furthermore, the annealing process itself can lead to an interdiffusion at the substrate interface,\cite{mitra_interfacial_2017, cooper_unexpected_2017} often leading to the formation of a magnetic dead layer, \cite{mitra_interfacial_2017, cooper_unexpected_2017, will-cole_negligible_2023} as well as an increase of the ferromagnetic resonance linewidth, especially at low temperatures. \cite{haidar_thickness-_2015, jermain_increased_2017} On the one hand, YIG grown on GGG by LPE requires no post annealing, which allows for the suppression of the gadolinium interdiffusion, leading to an extremely sharp interface.\cite{will-cole_negligible_2023} On the other hand, scaling the LPE process is not straightforward. Sputter deposition \cite{chang_nanometer-thick_2014} or solution based methods \cite{cao_van_thickness_2022,seol_development_2023} allow for wafer scale processes, but the mandatory post annealing step should be optimized to allow fast processing, which then could simultaneously reduce the interdiffusion of yttrium and gadolinium. To achieve this, the annealing time required to yield fully crystalline YIG films needs to be kept as low as possible.}
		
	\green{However, the dynamics describing the crystallization of YIG thin films during the post annealing step are only selectively reported in the literature. Typically, only the temperature and a time proven to yield a completely crystalline thin film with the desired properties are reported.}
	
	Here, we present an extended picture of the crystallization dynamics of YIG at different temperatures and annealing times, which allows us to define different crystallization windows depending on the substrate material. Our results provide a general crystallographic description of the crystallization process for YIG thin films and summarize the crystallization parameters found in the literature.
	
	\section{Methods}
	Ahead of the deposition, all substrates were cleaned for five minutes in aceton and isopropanol, and one minute in de-ionized water in an ultrasonic bath. YIG thin films were then deposited \green{at room temperature} onto different substrate materials using RF sputtering from a YIG sinter target at \SI{2.7}{\cdot 10^{-3}\,mbar} argon pressure and \SI{80}{W} power, at a rate of \SI{0.0135}{nm/s}. The nominal thickness upon deposition was \SI{33}{nm}. The post-annealing steps were carried out in a tube zone furnace under air. 
	
	As substrates yttrium aluminum garnet (\ch{Y3Al5O12}, YAG, \textit{CrysTec}) and gadolinium gallium garnet (\ch{Gd3Ga5O12}, GGG, \textit{SurfaceNet}) with a <111> crystal orientation along the surface normal were used. Additionally, silicon wafers cut along the <$100$> crystal direction with a \SI{500}{nm} thick thermal oxide layer (\ch{Si}/\ch{SiO_x}, \textit{MicroChemicals}) were used. Since GGG and YAG crystallize in the same space group Ia$\bar{3}$d as YIG and their lattice parameters are \SI{1.2376}{nm} \cite{Gates_Rector_2019_GGG} and \SI{1.2009}{nm}, \cite{Gates_Rector_2019_YAG} respectively, they are considered lattice matched in regards to the \SI{1.2380}{nm} for YIG. \cite{Gates_Rector_2019_YIG} The lattice mismatch $\epsilon$ can be calculated with Eq. \eqref{eq_lm} 
	\begin{equation}
		\epsilon = \frac{a_{YIG}-a_{substrate}}{a_{substrate}} \cdot 100\%
		\label{eq_lm}
	\end{equation}
	and translates to \SI{0.03}{\%} for GGG and \SI{3.09}{\%} for YAG. \cite{gross_festkorperphysik_2014} Due to the amorphous \ch{SiO_x} layer the \ch{Si}/\ch{SiO_x} substrates do not provide any preferential direction for crystallization. But even considering the underlying Si layer, we do not expect it to influence the crystallization direction in any way, as it features a fundamentally different space group (Fd$\bar{3}$m) and lattice constant. \cite{Gates_Rector_2019_Si} Therefore, \ch{Si}/\ch{SiO_x} is considered non lattice matched and fulfills the function as an arbitrary substrate.
	
	For the structural characterization X-ray diffraction measurements (XRD) were performed using a Rigaku Smart Lab Diffractometer with Cu $K_{\alpha}$ radiation. Scanning electron microscopy as well as electron backscatter diffraction (EBSD) measurements were conducted using a Zeiss Gemini Scanning Electron Microscope (SEM). \green{The magnetic properties were characterized via magneto-optical Kerr effect measurements in longitudinal geometry (L-MOKE) in a commercial Kerr microscope from Evico Magnetics.}
	
	\section{Results and Discussion}	
	\begin{figure}[b]
		\begin{center}
			\includegraphics[width=\linewidth]{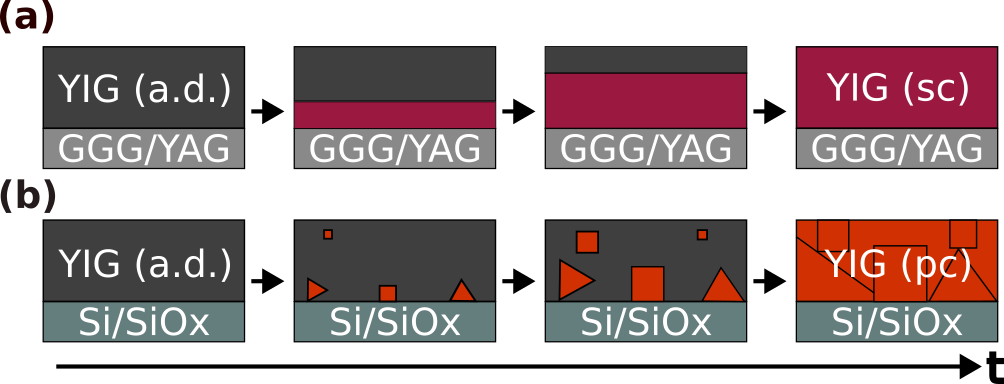}
			\caption{Expected crystallization of an amorphous, as-deposited (a.d.) YIG thin film on lattice matched substrates (a) and non lattice matched substrates (b). In the first case of solid phase epitaxy, a homogeneous crystal front forms at the substrate and propagates towards the upper thin film border. For the latter, nucleation is necessary and crystallites form in various orientations. This results in a single crystalline (sc) film for the epitaxy and a polycrystalline (pc) film when nucleation occurs.}
			\label{Fig_1}
		\end{center}
	\end{figure}
	
	\begin{figure*}[t]
		\begin{center}
			\includegraphics[width=\linewidth]{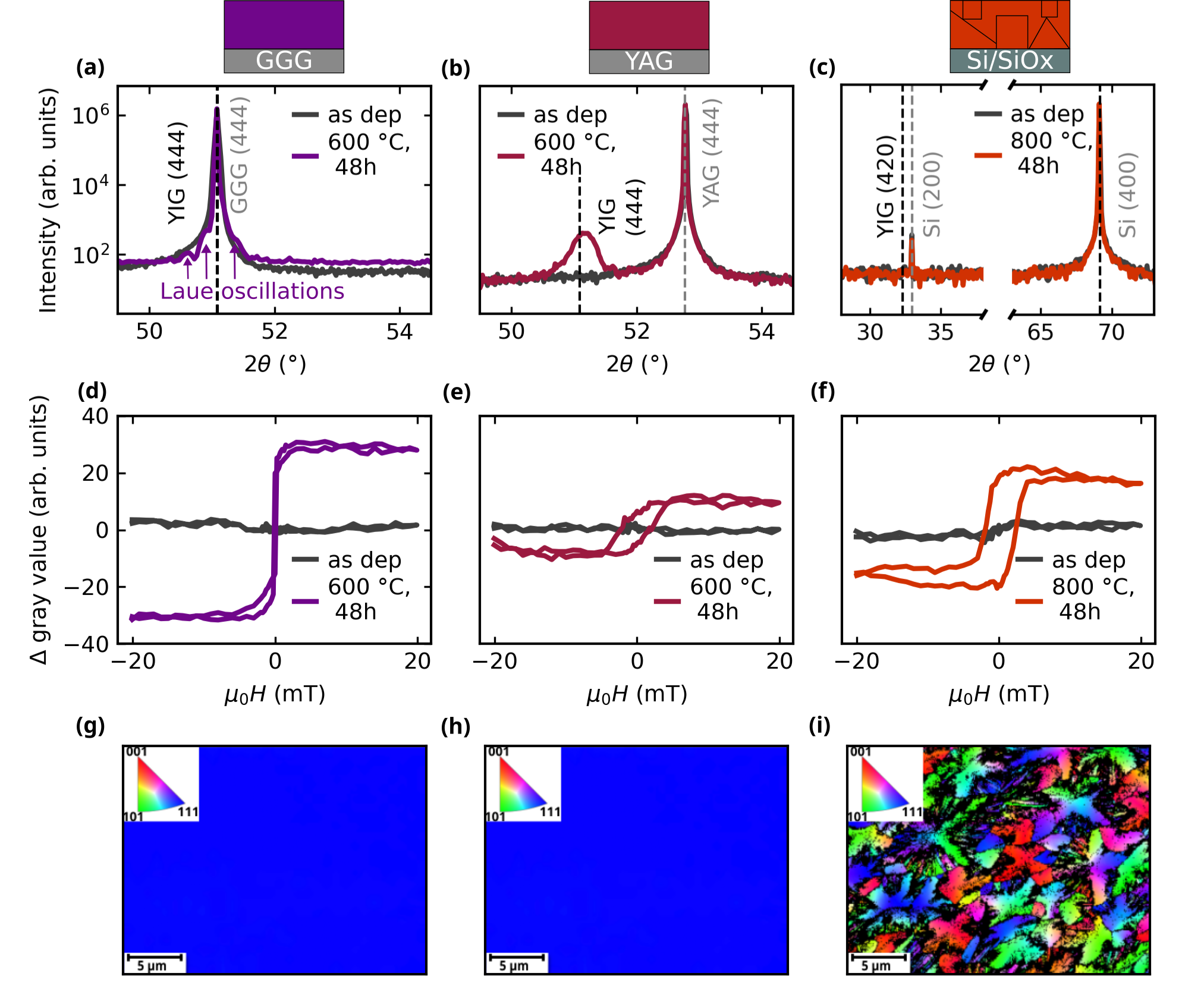}
			\caption{(a)-(c) XRD analysis of YIG thin films pre and post annealing on different substrates as given above in the respective column. The nominal positions of the substrate and the thin film are marked by the grey and black dashed lines, respectively. The additional peak marked with Si(200) in (c) is a detour reflex from the substrate. \green{(d)-(f) Background corrected Kerr microscopy data in L-MOKE configuration for the same samples before and after the annealing procedure. The change in the measured gray value corresponds to a change in the magnetization of the sample. The data was acquired from a central spot on the sample.} (g)-(i) crystal orientation of the post annealed YIG thin films normal to the surface normal as extracted from the Kikuchi-patterns determined by EBSD. The as-deposited films showed no Kikuchi-Patterns and are therefore not shown here.}
			\label{Fig_2}
		\end{center}
	\end{figure*}
	The crystallization mechanism of a thin film crucially depends on the substrate: for substrates where the lattices of film and substrate are sufficiently similar, the thin film layer crystallizes epitaxially, whereas for a substrate where the two lattices do not match, nucleation is needed. 
	
	Figure~\ref{Fig_1} shows the different crystallization mechanisms and the resulting YIG micro structure depending on the chosen substrate. As depicted in Figure~\ref{Fig_1}(a), a lattice matched substrate acts as a seed on which the film can grow epitaxially. Therefore, a single crystalline front is expected to move from the substrate towards the upper boundary of the film, \cite{csepregi_chaneling_1975, csepregi_regrowth_1976} which is commonly referred to as solid phase epitaxy (SPE) in the literature. For a substrate with a sufficiently large lattice mismatch or no crystalline order, no such interface is given, see Fig.~\ref{Fig_1}(b). Here, a nucleus needs to be formed first from which further crystallization takes place. The formation of the initial seeds by nucleation is expected to yield random crystal orientations. The polycrystalline seeds grow until reaching another grain or one of the sample's boundaries. For any of these processes, SPE or nucleation, to take place, the system needs to be at a temperature characteristic for this specific thin film/substrate system. \cite{chen_distinct_2017}
	
	To distinguish between amorphous, partly and fully crystalline films we apply several characterization methods, probing the structural and magnetic properties of the YIG thin films. The typical fingerprints of amorphous versus crystalline YIG on different substrates as determined by X-ray diffraction (XRD), \green{the longitudinal magneto-optical Kerr effect (L-MOKE)} and electron back scatter diffraction (EBSD) are depicted in Figure~\ref{Fig_2}. From top to bottom we gain an increased spacial resolution, probing increasingly smaller areas of the sample. 
	
	With XRD, the structural properties of YIG on YAG and GGG can be evaluated. For the amorphous films, the XRD measurements in Fig.~\ref{Fig_2} (a-c) show a signal stemming only from the substrate (cp. gray dashed lines). Upon annealing, YIG is visible in the form of Laue-oscillations on GGG (purple) and as a peak on YAG (red). In stark contrast to that no signal, which could be attributed to YIG, can be found on \ch{SiO_x}, even when annealing at \SI{800}{\degreeCelsius} for \SI{48}{h}. The sharp peak in Fig.~\ref{Fig_2}(c) at \SI{32.96}{\degree} can be attributed to a detour excitation of the substrate, as it is visible in the as deposited state and fits the forbidden Si (200) peak. \cite{renninger_umweganregung_1937} In the literature, YIG on \ch{SiO_{x}} has been reported to be polycrystalline at lower annealing temperatures than in the exemplary data shown in Fig.~\ref{Fig_2}(c). \cite{boudiar_magneto-optical_2004, kang_magnetic_2005, roumie_effect_2010} These films show peaks in the XRD, however they were at least one order of magnitude thicker. We therefore do not expect the YIG layer on \ch{Si}/\ch{SiO_x} to be amorphous, which will be confirmed in the following. 
	
	By probing the magnetic properties of the thin films with \green{L-MOKE} (cp.~Fig.~\ref{Fig_2}(d-f)), a clear distinction between amorphous and crystalline YIG can be made. While the film shows a linear \green{L-MOKE} signal in the as-deposited state, it changes to a hysteresis for all three samples upon annealing. In general, the sharpest hysteresis is visible for YIG on GGG, which becomes broader for an increasing structural misfit. Na\"{i}vely polycrystalline samples are expected to consist of multiple domains pointing towards different directions, which lead to an increase of the coercive field. This is consistent with our results and also with the magnetic properties found in literature. \cite{kang_magnetic_2005, zheng_preparation_2014, li_epitaxial_2016, hauser_yttrium_2016} These coercive fields are below \SI{0.1}{mT} for YIG on GGG\cite{li_epitaxial_2016, hauser_yttrium_2016} and between 2.2-\SI{3}{mT} for YIG on \ch{Si}/\ch{SiO_x}.\cite{kang_magnetic_2005, zheng_preparation_2014} The \green{L-MOKE} measurements therefore indicate the spontaneous formation of a phase with magnetic ordering on all three substrates.
	
	\green{For additional characterization of the magnetic properties of the films via ferromagnetic resonance and SQUID magnetometry please refer to the supplemental information.\cite{SI_Sailler_2024} The corresponding data show the same dependence on the type of substrate, that is also apparent in the L-MOKE measurements. Once the YIG is fully crystallized, however, we do not find a dependence of the magnetic parameters of our thin films on the annealing parameters.}
	
	While L-MOKE correlates the magnetic properties with amorphous and crystalline films, it lacks the ability to quantify the amount of crystalline YIG. The hysteretic response for the annealed YIG on \ch{SiO_x} strongly supports the formation of crystalline YIG, however, we cannot correlate this to a percentage of crystalline material. Therefore, a structural characterization with higher spacial resolution than XRD is needed.
	
	To that end electron back scatter diffraction (EBSD) measurements were performed. With this technique Kikuchi patterns, which are correlated to the crystal structure, are detected and later evaluated. The results are shown for crystalline samples only, as the amorphous film showed no Kikuchi patterns. This confirms, that the detected patterns stem from the YIG thin film itself and not from the crystallographically similar substrates of YAG or GGG. This is consistent with the EBSD signal depth given in the literature of 10 to \SI{40}{nm}. \cite{dingley_progressive} The extracted crystal orientations along the surface normal can be seen in Fig.~\ref{Fig_2} (g-i). On YAG and GGG a single color corresponding to the <$111$> direction is visible in the mapping, which is consistent with the XRD data and corroborates SPE from the substrate in the <$111$> direction. On \ch{SiO_{x}}, however, various crystal directions are present, confirming the polycrystalline nature of the YIG. The crystallographic data from our EBSD measurements show random nucleation. The cross shape of the individual crystalline areas point towards an anisotropic crystallization with a preferential direction along <$110$> or higher indexed directions like <$112$>, which is consistent with earlier studies on YIG and other rare earth garnets \cite{nielsen_growth_1958, tolksdorf_facet_1981, bennema_morphology_1983, park_structural_2001} as well as PLD grown bismuth iron garnet. \cite{heinrich2006pulsed}
	
	\green{The use of EBSD enables the quantification of the amount of crystalline material in a YIG thin film on \ch{SiO_x} or any arbitrary substrate. Combining the magnetic and structural data from L-MOKE and EBSD, respectively, allows for an unambiguous identification of the formation of polycrystalline YIG on \ch{SiO_x}.
		
	We presume that the absence of any XRD peaks in the symmetric $\theta - 2\theta$ scan results from the small volume of the individual crystallites of YIG on \ch{SiO_x}. \cite{zhang_growth_2015, lian_annealing_2017, cao_van_thickness_2022}} We approximate the volume of a single polycrystalline grain, i.e. one cross from the EBSD data (cp. Fig.~\ref{Fig_2}(i)) to be \SI{0.5}{\mu m^3}, stemming from an area of about \SI{15}{\mu m²} and a film thickness of \SI{32}{nm}. This is also the size of individually contributing grains to the diffraction within the XRD. Assuming a single crystalline thin film, where the whole irradiated area contributes additively, the contributing area amounts to \SI{7}{\cdot10^5 \mu m^3} , which is six orders of magnitude larger than that of an individual grain. Therefore, the contributions of the individual grains of the YIG layer on \ch{SiO_x} to the XRD intensity are too small to result in a finite peak for a \SI{30}{nm} thick film. 
	
	\begin{figure}[h]
		\begin{center}
			\includegraphics[width=\linewidth]{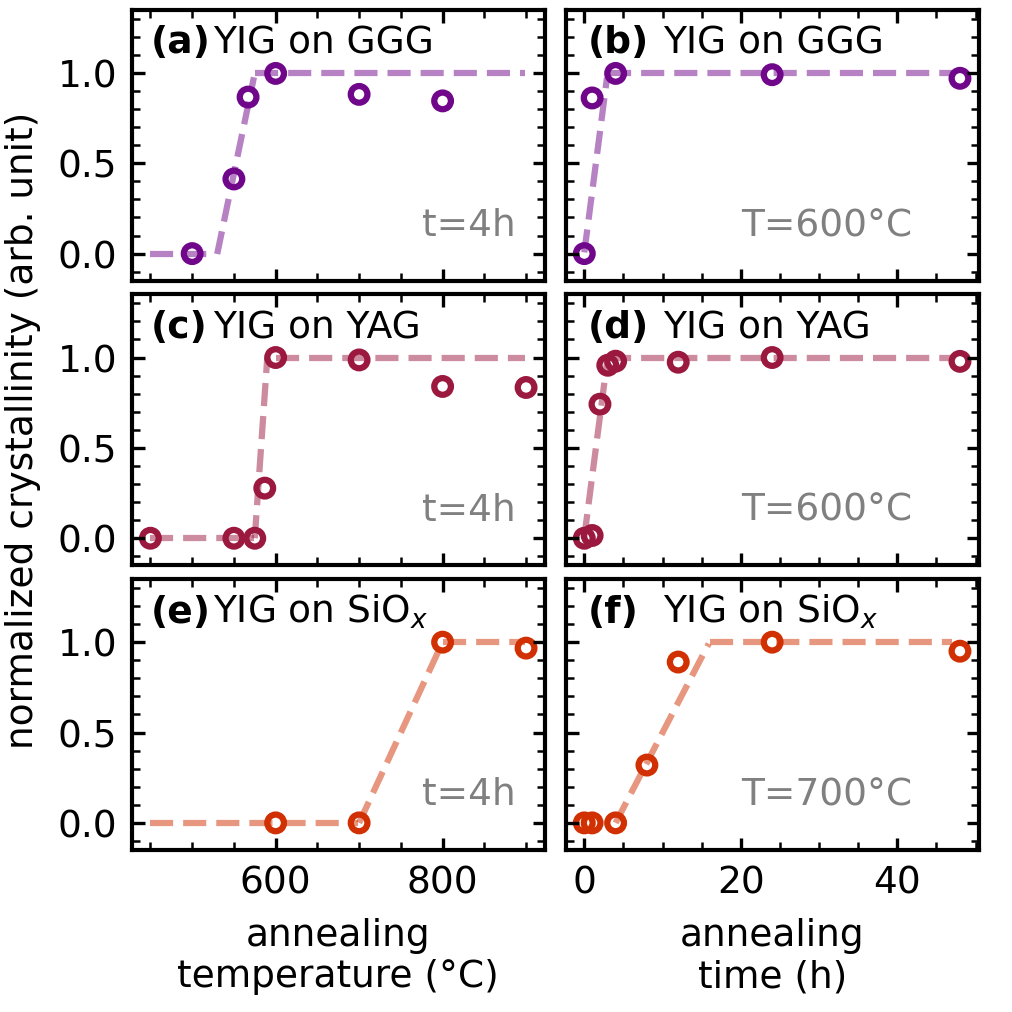}
			\caption{Evolution of the crystallinity as a function of the annealing temperature for a constant annealing time of \SI{4}{h} (a,c,e) and for different times at a constant temperature of \SI{600}{\degreeCelsius} on GGG, YAG (b,d) or \SI{800}{\degreeCelsius} on \ch{SiO_x}(f). The dotted lines act as a guide to the eye.}
			\label{Fig_3}
		\end{center}
	\end{figure}
	
	These results provide the basis for the investigation of the crystallization behavior and reveal how different techniques enable us to distinguish between amorphous, partly and fully crystalline films. We utilize the structural information to analyze the crystallization dynamics on the different substrates.
	
	The percentage of crystalline YIG was quantified differently for the three different substrates. For YIG on YAG the amount of crystalline YIG correlates to the intensity of the Bragg peak. A certain film thickness corresponds to a maximum area under the peak, to which the intensity is normalized. For YIG on GGG, the percentage of crystalline YIG is extracted from the Laue oscillations (cp. Fig~\ref{Fig_2}(a)). The frequency of the oscillation corresponds to the number of interfering lattice planes, enabling the calculation of the thickness of the crystalline layer. Using X-ray reflectivity, the absolute film thickness was measured for each film. \green{For these measurements and evaluation, please refer to the supplemental information.\cite{SI_Sailler_2024}} Comparing the thickness of the crystalline layer with the film thickness then enables to monitor the crystallization of YIG on GGG. For the non lattice matched substrates, EBSD mappings were taken to extract the amount of crystalline YIG. \green{Further evaluation of partly crystalline YIG on \ch{SiO_x} can be found in the supplemental information.\cite{SI_Sailler_2024}} For each of the YIG thin films, a percentage of crystalline YIG at a given time and temperature is extracted, which allows an evaluation of the crystallization process for this specific temperature. 
	
	First, we find the onset temperature for the crystallization of YIG on each substrate. As crystallization is thermally activated, it depends exponentially on the annealing temperature, \cite{avrami_kinetics_1939} which leads to a very narrow temperature window of incomplete crystallization. To extract this window, multiple samples were annealed for four hours at different temperatures. \green{Figure~\ref{Fig_3} shows the results for YIG on GGG (a), YIG on YAG (c) and YIG on \ch{SiO_x} (e). At substrate dependent temperatures of \SI{550}{\degreeCelsius}, \SI{575}{\degreeCelsius} and \SI{700}{\degreeCelsius} for YIG on GGG (a), YAG (c) and \ch{SiO_x} (e), respectively, a steep increase in the crystallinity can be seen. Towards higher temperatures the extracted value stays the same or is only slightly reduced, which suggests, that the YIG film is fully crystallized and no further changes are expected. A crystalline YIG film on YAG and GGG can therefore be obtained at a temperature range around \SI{600}{\degreeCelsius}, whereas on \ch{SiO_x}, temperatures of approximately \SI{700}{\degreeCelsius} are necessary.}
	
	For our samples, the heating up and cooling down is included in the annealing time. An in-situ study on a representative sample with $d_{\mathrm{YIG}}=\SI{100}{\nano\meter}$ yielded data in good agreement with the crystallization behavior in the one zone furnace. It should be noted that the use of different equipment led to a small variation in the absolute temperature, see supplemental information.\cite{SI_Sailler_2024} 
	
	\green{The lower extracted crystallinities for YIG on YAG and GGG at \SI{800}{\degreeCelsius} and above (cp. Fig.~\ref{Fig_3}(a)+(c)) hint towards the occurrence of competing crystallization processes.} We attribute the reduction in crystallinity at annealing temperatures above \SI{800}{\degreeCelsius} to additional formation of polycrystals enabled by the elevated temperatures, which competes with the solid phase epitaxy and by that reduces the crystal quality of the thin film. Analyzing the rocking curves of these samples (see supplemental information) confirms an increased full width at half maximum value at higher temperatures.\cite{SI_Sailler_2024} This can be correlated with a lower crystal quality, which supports an additional crystallization process.
	
	\green{To study the crystallization dynamics, the time evolution of the normalized crystallinity for a given temperature is evaluated, shown in Fig.~\ref{Fig_3} for YIG on GGG(b), YAG(d) and \ch{SiO_x}(f). Here, a sample was subjected to the same temperature for multiple repeats until the extracted value and therewith the crystalline amount did saturate. This saturation can be seen on all substrates and represents a fully crystallized thin film, where no further changes are expected.}
	
	To describe the crystallization at an arbitrary temperature, we find a general crystallographic description for each of the substrates. A phase transition in a solid like crystallization can generally be described by the Avrami equation: \cite{avrami_kinetics_1939, william1939reaction, avrami_kinetics_1940, avrami_granulation_1941} 
	\begin{equation}
		\theta_c = 1 - e^{-k\cdot t^n}
		\label{eq_avr}
	\end{equation}
	where $\theta_c$ is the crystallinity normalized to one, with respect to a complete crystallization, $k$ the rate constant and $t$ the annealing time. The exponent $n$ is often referred to as the Avrami exponent and describes how the crystallization takes place.\cite{avrami_kinetics_1940} It can take values between 1 and 4, where one contribution stems from the nucleation and takes values of 0 for controlled and 1 for random nucleation, while the other contributions originate from the type of crystallization in the three spacial directions.
	\begin{figure}[b]
		\begin{center}
			\includegraphics[width=\linewidth]{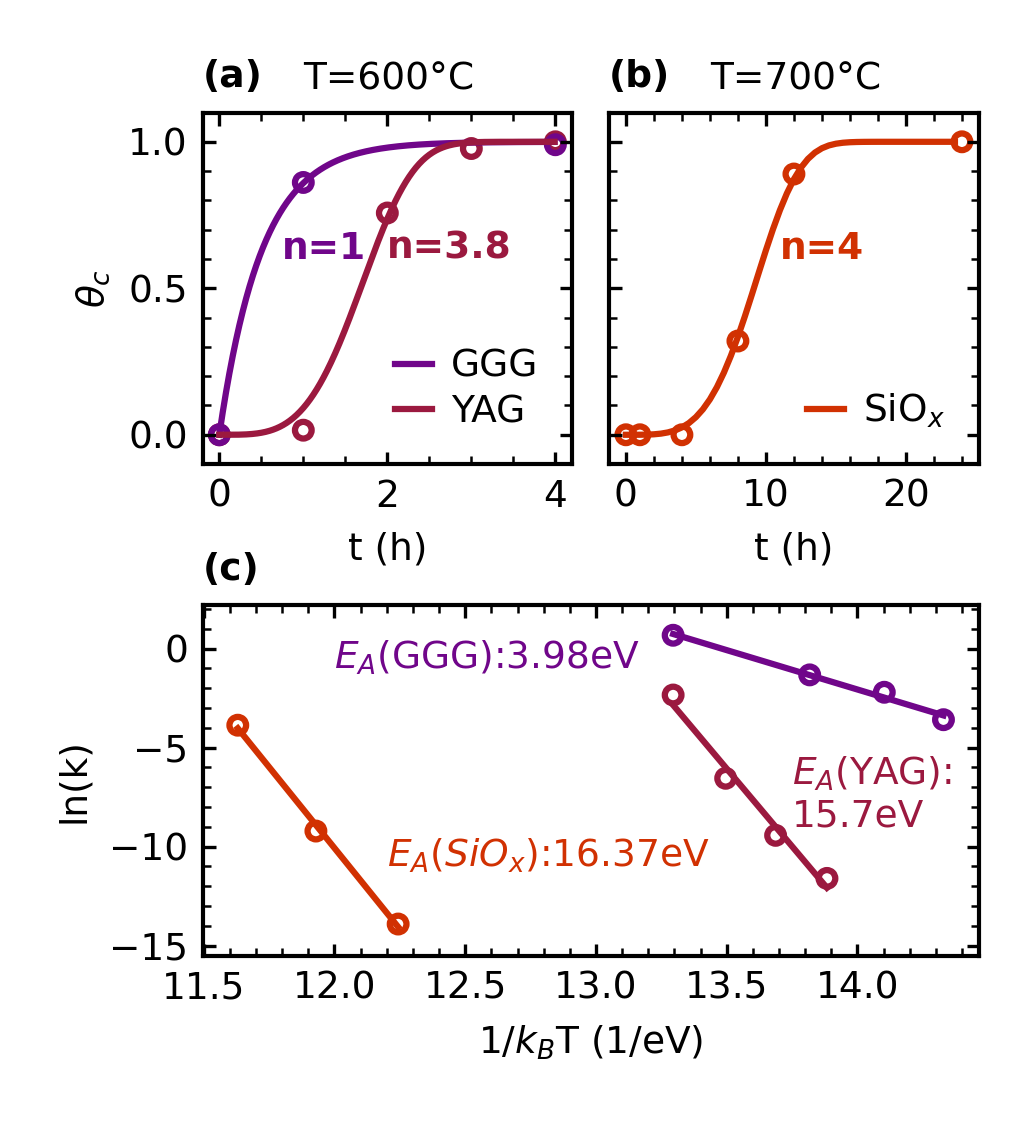}
			\caption{(a) and (b) show the time evolution of the YIG crystallization on the three substrates after normalizing the data with the maximum value to 1. The dots represent the crystallinity values from XRD (YAG/GGG) and EBSD (\ch{SiO_x}), while the solid lines show the fit of the data using Eq.\eqref{eq_avr}. Because of the inherently different crystallization processes, the time scales and the temperatures differ. Conducting these time evolutions at different temperatures for each substrate result in a rate constant $k(T)$ for this temperature. A logarithmic representation of the $k(T)$ values over the inverse temperature is given by the symbols in (c). For each substrate a linear expression was fitted, where the slope represents the activation Energy $E_A$ and the intercept of the y-axis the pre-factor $k_0$ for YIG on each substrate.}
			\label{Fig_4}
		\end{center}
	\end{figure}
	\begin{figure*}[t]
		\begin{center}
			\includegraphics[width=\linewidth]{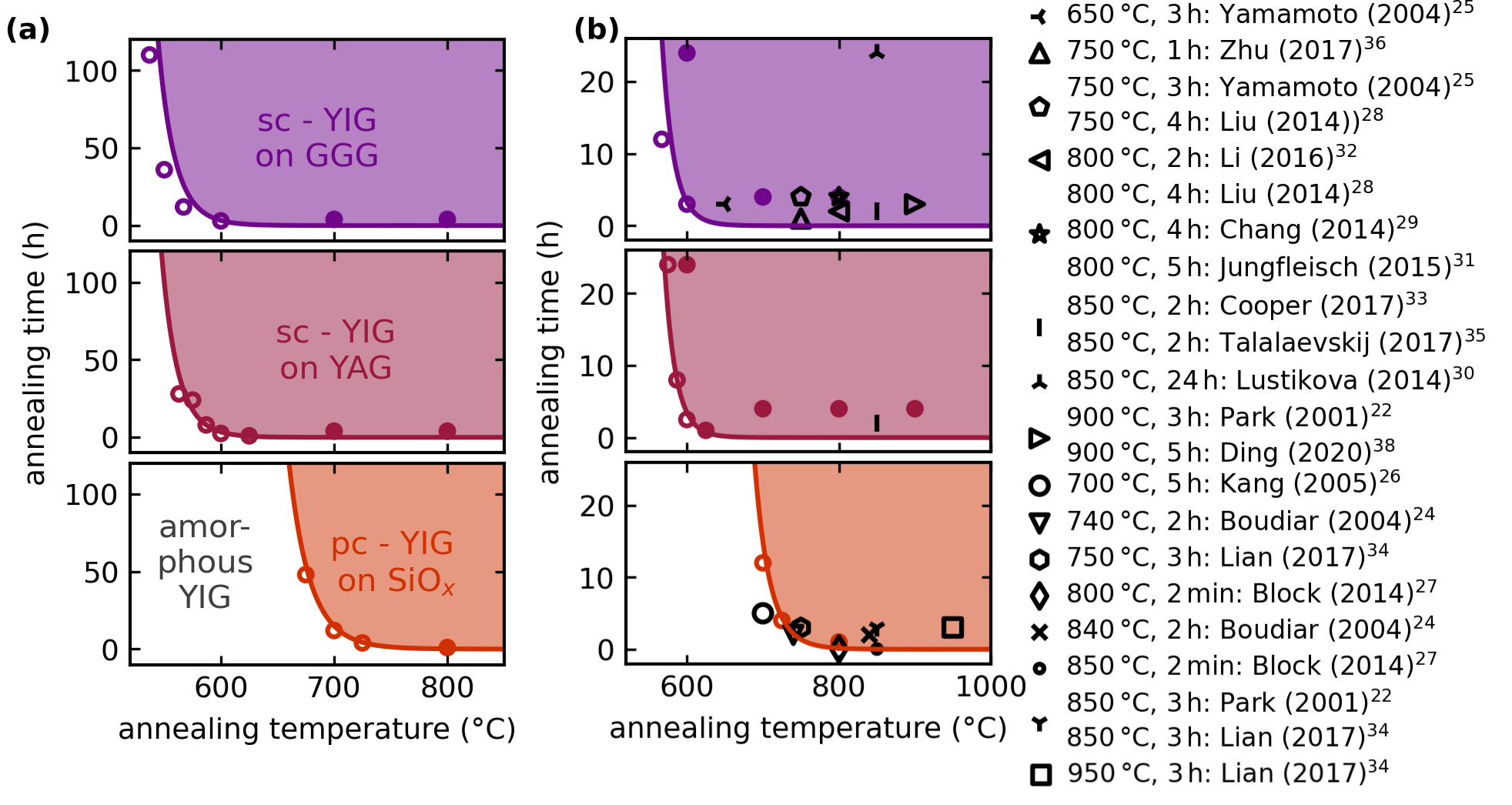}
			\caption{(a) Annealing parameters to obtain a fully crystalline YIG film on the respective substrates. We expect every point in the colored area to yield a fully crystalline sample. We use Eq.\eqref{t_avr} with the values obtained in Fig.~\ref{Fig_4}(c) to determine the boundary separating crystalline YIG (shaded areas, sc = single crystalline, pc = poly crystalline) from amorphous YIG (white areas). The open circles represent the samples from Fig.~\ref{Fig_4}(c) which are used for the fit. Further studied, fully crystalline samples are marked by the full circles. There are different regions where the YIG is fully crystalline depending on the substrate. Panel (b) gives a comparison of our crystallization diagram with other studies. \cite{park_structural_2001, jang_new_2001,boudiar_magneto-optical_2004, yamamoto_post-annealing_2004, kang_magnetic_2005, block_growth_2014, liu_ferromagnetic_2014, chang_nanometer-thick_2014, lustikova_spin_2014, jungfleisch_spin_2015, li_epitaxial_2016, cooper_unexpected_2017, lian_annealing_2017, talalaevskij_magnetic_2017, zhu_patterned_2017, ding_nanometer-thick_2020, ding_sputtering_2020} \green{Note that, while we here consider only the crystallization of sputtered thin films by post annealing, the crystallization diagram also fits for comparable samples obtained by PLD (not shown here). \cite{heinrich2006pulsed, krockenberger_solid_2008, haidar_thickness-_2015, hauser_yttrium_2016, hauser_annealing_2017, heyroth_monocrystalline_2019, gurjar_control_2021}}}
			\label{Fig_5}
		\end{center}
	\end{figure*}
	For the rate constant k we use an exponential Arrhenius dependency:\cite{callister_materials_2007, csepregi_chaneling_1975}
	\begin{equation}
		k = k_0 \cdot e^{\frac{-E_A}{k_B\cdot T}}
		\label{eq_arr}
	\end{equation}
	where both the pre-factor $k_0$ and the activation energy $E_A$ are unique for each combination of film and substrate material.
	
	
	The Avrami equation (cp.~Eq.\eqref{eq_avr}) lets us describe the crystallization on all three substrates. To that end we fit the normalized crystallinity values of YIG with the Avrami equation (cp.~Eq.\eqref{eq_avr}), where we fix the Avrami exponent n between $1$ and $4$ (cp. Fig.~\ref{Fig_4}(a)+(b)). The rate constants $k$ then describe the crystallization velocities on the respective substrate in $h^{-1}$. The crystallization behavior of YIG on GGG and YAG at an annealing temperature of \SI{600}{\degreeCelsius} is shown in Fig.~\ref{Fig_4}(a). 
	
	On GGG at \SI{600}{\degreeCelsius} (cp. Fig.~\ref{Fig_4}(a)), YIG immediately starts to crystallize with a rate constant of \SI{1.96}{h^{-1}} and an Avrami exponent of 1. This means, that the crystallization takes place without nucleation and in one spacial direction, which is consistent with the monotonously moving crystallization front expected for SPE. The rate constant translates to a initial velocity of \green{\SI{0.98}{nm/min}} for the \SI{30}{nm} films. Towards longer annealing times, the curve flattens, meaning that the crystalline material reaches the sample's surface.
	
	The crystallization of YIG on YAG shows an initial time delay, despite the comparably small lattice mismatch of \SI{3.09}{\%} (cp. Fig.~\ref{Fig_4}(a)). The fitting of the data at \SI{600}{\degreeCelsius} leads to a rate constant of \SI{0.10}{h^{-1}} with $n = 3.8$. This means, that the crystallization does not follow a typical SPE behavior and nucleation processes in the thin film cannot be excluded. However, also for the crystallization on YAG, single crystalline YIG is obtained (cp. Fig.~\ref{Fig_2}(b)+(h)). This deviation from YIG on GGG is most likely due to the larger lattice mismatch which causes an energetically costly strain in the film. \cite{wang_strain-tunable_2014} The crystallization velocity along the surface normal direction is obtained by the $n$-th root out of the rate constant and translates to \SI{0.27}{nm/min}. 
	
	The crystallization of YIG on \ch{SiO_{x}} is fundamentally different (cp. Fig.~\ref{Fig_4}(b)). Here, polycrystalline grains were found at temperatures of \SI{675}{\degreeCelsius} and above. The time evolution of the crystallinity is depicted in Figure~\ref{Fig_4}(b), where fitting the data by the Avrami equation (Eq. \eqref{eq_avr}) yields $n = 4$ and a rate constant of \SI{9.9}{\cdot 10^{-5}\,h^{-1}}. This confirms our initial hypothesis of nucleation and subsequent crystallization in three dimensions. Higher temperatures compared to the garnet substrates are needed to provide enough energy for nucleation, which causes the crystallization process to be visible at \SI{675}{\degreeCelsius} and above. 
	
	An approximation of the crystallization velocity can be extracted from the EBSD data. Here, we assume that the crystallization starts in the middle of a cross shape structure (cp. Fig.~\ref{Fig_2}(i)) and stops when reaching a boundary given by neighboring crystallites. The distance covered depends on the number of nuclei formed and is highly dependent on the crystallographic direction. To ensure comparability with the two lattice matched substrates, we consider grains growing in plane along the <$111$> direction. At \SI{700}{\degreeCelsius}, the YIG crystallites on \ch{SiO_{x}} measured up to \SI{10}{\mu m} in length after at least \SI{12}{h} of annealing. This translates into a propagation velocity of \SI{16.7}{nm/min} at \SI{700}{\degreeCelsius} on an arbitrary substrate along the <$111$> direction. 
	
	
	To compare the three crystallization velocities, the temperature dependence of the rate constants $k$ needs to be taken into consideration. Using the Arrhenius equation (Eq. \eqref{eq_arr}) we are able to extrapolate the crystallization rate at any temperature. To that end, the logarithm of each rate constant is plotted over the inverse temperature. The linear dependency of Eq. \eqref{eq_arr} in the logarithmic plot allows us to extract the activation energy and the pre-factor $k_0$ for YIG on each substrate. The resulting values are plotted in Tab. \ref{tab_1}. While at first glance the crystallization velocity for YIG on \ch{SiO_x} seems faster, the different annealing temperatures of \SI{600}{\degreeCelsius} for the garnet substrates and \SI{700}{\degreeCelsius} for \ch{SiO_x} need to be taken into account (cp.~Fig.\ref{Fig_4}). Extrapolating the crystallization velocity for YIG on GGG at \SI{700}{\degreeCelsius} reveals that here YIG would crystallize approximately 30 times faster than on \ch{SiO_x}.
		
	Our activation energy of \green{\SI{3.98}{eV}} for YIG on GGG is in good agreement with the literature. For the formation of bulk YIG from oxide powders, a value of \SI{5.08}{eV} was reported. \cite{wan_ali_investigation_2016} Further, for the crystallization of bulk polycrystalline YAG, which is expected to behave similarly as it has the same crystal structure, an activation energy of \SI{4.5}{eV} was found.\cite{johnson_crystallization_2001} The lower value of \green{\SI{3.98}{eV}} for YIG on GGG highlights the reduced energy needed, due to the SPE from the lattice matched GGG.
	
	The activation energies for YIG on YAG as well as on \ch{SiO_x} are much higher than the value on GGG. As the general crystallization windows and times needed for a fully crystalline film stay the same, we ascribe this behavior to a kinetic blocking, originating from the lattice mismatch and the nucleation. Understanding the exact mechanism however, would need further study.
	
	These results allow to establish a diagram to underline which annealing parameters will lead to a fully crystalline YIG thin film on the three substrates (cp. Fig.~\ref{Fig_5}(a)). For a mathematical description, we combine the Avrami equation Eq. \eqref{eq_avr} with the Arrhenius equation Eq. \eqref{eq_arr} to be able to express the crystallinity in terms of annealing time and temperature.
	\begin{equation}
		t = \left(\left[-\frac{ln(1-\theta_c)}{k_0}\right]\cdot e^{\frac{E_A}{k_B T}}\right)^{\frac{1}{n}}
		\label{t_avr}
	\end{equation}
	We use a crystallinity $\theta_c$ of 0.999 to avoid the divergence of the logarithm and the respective $n$, $k_0$ and $E_A$ found in Tab.~\ref{tab_1}.
	
	\begin{table}[b]
		\centering
		\newcolumntype{Y}{>{\centering\arraybackslash}X}
		\caption{Extracted activation energies $E_A$ and pre-factors $k_0$ for YIG on each substrate}
		\begin{tabularx}{\linewidth}{ Y  Y  Y Y }
			\hline \hline
			& $E_A$ (eV) & $k_0$ ($1/\,\mathrm{h}$) & n \\ \toprule
			{YIG on GGG} & {$3.98 \,\pm$ 0.32} & { \SI{2.0}{\cdot10^{23}} } & 1 \\ \midrule
			YIG on YAG & $15.70\, \pm$ 1.59 & \SI{2.6}{\cdot 10^{89}} & 3.8 \\ \midrule
			YIG on \ch{SiO_x} & $16.37\, \pm$ 0.85 & \SI{8.4}{\cdot 10^{80}} & 4\\ \hline \hline \label{tab_1}
		\end{tabularx}
	\end{table}
	
	Figure~\ref{Fig_5}(a) outlines the temperature and time combination where crystalline YIG (shaded areas) can be obtained. Regions where the YIG thin film remains amorphous are left in white. The boundary between non crystalline and crystalline for each substrate is given by Eq. \eqref{t_avr}. Each of the circles seen in Fig.~\ref{Fig_5}(a) represents one fully crystalline sample obtained as described for Fig.~\ref{Fig_3}(b). The filled circles represent fully crystalline samples, where no time dependence of the crystallinity was measured. As already anticipated, YIG exhibits different crystallization behavior depending on the substrate. Note, that the formation of polycrystalline YIG on \ch{SiO_{x}} or any arbitrary substrate needs notably higher temperatures than SPE, where an annealing at \SI{660}{\degreeCelsius} for \SI{100}{h} would be necessary to result in a fully crystalline film. 
	
	The different temperatures and times necessary to induce crystallization stem from the different types of substrates. For YIG on GGG and YAG the seed for the crystallization is given by the lattice of the substrate. Therefore, we ascribe the discrepancy between YAG and GGG to the different lattice mismatch compared to YIG. In the YIG thin films on YAG a higher strain is expected to exist in the film, which leads to the formation of energetically costly dislocations. This in turn results in the slightly higher temperature needed for YIG to crystallize on YAG. On \ch{SiO_{x}}, however, a significantly higher temperature than for the lattice matched substrates is needed for crystalline YIG to form. Here, as no initial seed is given by the substrate, nucleation is required, which is a thermally activated process that needs additional energy, i.e. higher temperatures. This random formation of seeds leads to a polycrystalline YIG thin film on \ch{SiO_x}
	
	A comparison with the literature shows, that parameters which have been previously reported to result in a fully crystalline YIG layer, fit into our extracted area, (cp. Fig.~\ref{Fig_5}(b)).\cite{park_structural_2001, jang_new_2001,boudiar_magneto-optical_2004, yamamoto_post-annealing_2004, kang_magnetic_2005, block_growth_2014, liu_ferromagnetic_2014, chang_nanometer-thick_2014, lustikova_spin_2014, jungfleisch_spin_2015, li_epitaxial_2016, cooper_unexpected_2017, lian_annealing_2017, talalaevskij_magnetic_2017, zhu_patterned_2017, ding_nanometer-thick_2020, ding_sputtering_2020} Additionally to the sputtered films, also amorphous films obtained from PLD with subsequent annealing fit in the observed regions.\green{\cite{heinrich2006pulsed, krockenberger_solid_2008, haidar_thickness-_2015, hauser_yttrium_2016, hauser_annealing_2017, heyroth_monocrystalline_2019, gurjar_control_2021}} The extracted diagram in Fig.~\ref{Fig_5} therefore acts as a general description for the crystallization of YIG thin films out of the amorphous phase. 
	
	\section{Conclusion}
	Extensive time and temperature series were used to analyze the crystallization kinetics of sputtered amorphous YIG thin films on different substrates. We find the formation of single crystalline YIG thin films on garnet substrates where the crystallization on gadolinium gallium garnet can be coherently described in a solid phase epitaxy picture, whereas a more complicated crystallization scheme is found on yttrium aluminum garnet. On \ch{SiO_x} a polycrystalline YIG thin film develops, with slower crystallization dynamics than for the garnet substrates.
	
	A fully crystalline YIG film on GGG was found for temperatures as low as \SI{537}{\degreeCelsius} and annealing times of \SI{110}{h}. On silicon oxide (representing any type of amorphous or non lattice matched substrate), the nucleation of the YIG crystals is not expected for reasonable time scales below \SI{660}{\degreeCelsius}. The results summarized in Tab.~\ref{tab_1} allow for the determination of the crystallization velocity of YIG on those substrates for any temperature. 
	
	Thus, we provide a complete description of the crystallization process from the amorphous phase for YIG on GGG, YAG and arbitrary substrates such as \ch{SiO_x}, which allows us to define the range in which crystalline YIG thin films can be obtained. 
	
	\section{Acknowledgments}
	This work was funded by the Deutsche Forschungsgemeinschaft (DFG, German Research Foundation) – Project-ID 446571927 \green{and via the SFB 1432 - Project-ID 425217212}. We cordially thank F. Michaud and J. Ben Youssef from the Université de Bretagne Occidentale in Brest (France) for fruitful discussions and for letting us use their in-situ X-ray diffractometer. We also gratefully acknowledge technical support and advice by the nano.lab facility of the University Konstanz. 
	
	\section{References}
	\bibliographystyle{apsrev4-1}
	\bibliography{PaperBib_rev.bib}
	
\end{document}